\documentclass[nojss]{jss}

\usepackage{thumbpdf,lmodern}
\usepackage{graphicx}
\usepackage{hyperref}
\usepackage{amsmath}
\usepackage{framed}
\usepackage{float}

\author{Izhar Asael Alonzo Matamoros \\ \hspace{6cm} Universidad de Nacional Autónoma de Honduras \And 
Cristian Andres Cruz Torres \\}
\Plainauthor{Izhar Asael Alonzo Matamoros}

\title{ \pkg{varstan}: An \proglang{R} package for Bayesian analysis of structured time series models with \proglang{Stan}}
\Plaintitle{varstan: An R package for Bayesian analysis of structured time series models with Stan }
\Shorttitle{\pkg{varstan}: Bayesian time series analysis with \proglang{Stan} in \proglang{R}}

\Abstract{
\pkg{varstan} is an \proglang{R} package \citep{R} for Bayesian analysis of time series models using \proglang{Stan} \citep{Stan}. The package offers a dynamic way to choose a model, define priors in a wide range of distributions, check model's fit, and forecast with the m-steps ahead predictive distribution. The users can widely choose between implemented models such as \textit{multiplicative seasonal ARIMA, dynamic regression, random walks, GARCH, dynamic harmonic regressions,VARMA, stochastic Volatility Models, and generalized t-student with unknown degree freedom GARCH models}. Every model constructor in \pkg{varstan} defines weakly informative priors, but prior specifications can be changed in a dynamic and flexible way, so the prior distributions reflect the parameter's initial beliefs. For model selection, the package offers the classical information criteria: AIC, AICc, BIC, DIC, Bayes factor. And more recent criteria such as Widely-applicable information criteria (\textit{WAIC}), and the Bayesian leave one out cross-validation (\textit{loo}). In addition, a Bayesian version for automatic order selection in seasonal ARIMA and dynamic regression models can be used as an initial step for the time series analysis.\\
}

\Keywords{Time series, Bayesian analysis, structured models, \proglang{Stan}, \proglang{R}}
\Plainkeywords{Time series, Bayesian analysis, structured models,Stan, R}

\Address{
  Izhar Asael Alonzo Matamoros\\
  Universidad Nacional Autónoma de Honduras\\
  Departmento de Estadística\\
  Escuela de Matemática, Facultad de Ciencias\\
  Blvd Suyapa, Universidad Nacional Autónoma de Honduras\\
  Tegucigalpa, Honduras\\
  E-mail: \email{asael.alonzo@gmail.com}\\
  URL: \url{https://asaelam.wixsite.com/asael697site}
}

\begin{document}

\section[Introduction: ]{Introduction: } \label{sec:intro}

Structured models such as ARIMA \cite{box1970}, GARCH \cite{engle1982} and \cite{Bollersev1986}, Random walks and VARMA models are widely used for time series analysis and  forecast. Several \proglang{R} packages \citep{R} such as \pkg{forecast} \citep{Rob2007} and \pkg{astsa} \citep{Stoffer2019}, have been developed for estimating the models with classic inferences methods. Although a Bayesian approach offers several advantages such as incorporating prior knowledge in parameters, estimating the posterior distribution in complex models is a hard task and an analytical solution may not be feasible. A Markov chain Monte Carlo (MCMC) approach using Gibbs sampler limits prior selection to be conjugate to the likelihood or Metropolis-Hasting algorithms struggles with slow convergence in high dimensional models. The No U-Turn Sampler \cite{hoffman14a} algorithm provided by \proglang{Stan} \citep{Stan}  offers a fast convergence, prior flexibility and its own programming language for modeling (\textit{for a further discussion of samplers and algorithms} \cite{brms2017}). The package \pkg{varstan}, is an R interface of \proglang{Stan}'s language for time series modeling, offering a wide range of models, priors choice and methods making Bayesian time series analysis feasible.\\

The aim of this article is to give a general overview of the package functionality. First,  general definitions of structured time series models and priors selection. Then, a discussion of the estimating process is given. Also, package's functionality is introduced; as well as a presentation of the most important functions. Finally, an analysis of the monthly live births in the U.S.A (1948 -1979) is given as an example of the package modeling process.

\section{Structured Time series models} \label{sec:models}

A time series model is just a sample of a stochastic process ($Y = \{Y_t\}_{t=1}^\infty$), 
where every observation $Y_t$ describes the random variable response in a particular 
time $t$. Let's say the process follows a location-scale model \cite{Miggon} with normal distribution  
$Y_t \sim N(\mu_t,\sigma_t)$, where the mean ($\mu_t$) and variance ($\sigma_t$) are
considered the location and scale parameters with a time dependency. In other
words, every observation can be  written as follows:
$$Y_t = \mu_t + \sigma_t \epsilon_t, \  \epsilon_t \sim N(0,1) \ iid.$$   

The basic \textbf{ARIMA} model proposed by \cite{box1970}, can be seen as a location-scale model, where the time dependency structure is for the location parameter.
\begin{equation}
(1-B)^d Y_i = \mu_0 +\sum_{i = 0}^p \phi_i(1-B)^d Y_{t-i} - \sum_{i= 1}^q \theta_i \epsilon_{t-i} +\sigma_0 \epsilon_t \label{eq:1}
\end{equation}
where the previous equation is written as:
$$(1-B)^d Y_t = \mu_t +\sigma_t \epsilon_t$$
notice that:
\begin{itemize}
	\item B is the  back-shift operator, where $B^dY_t = Y_{t-d}$;
	\item d is the number of differences needed so the process is stationary;
	\item p is the number of considered lags in the auto-regressive component;
	\item q is the number of considered lags in the mean average component;
	\item $\mu_t =\mu_0 +\sum_{i = 0}^p \phi_i(1-B)^d Y_{t-i} - \sum_{i= 1}^q \theta_i \epsilon_{t-i}$ clearly has a time dependency;
	\item $\sigma_t  =\sigma_0$  does not have a time dependency because is constant in time; 
	\item $\epsilon_t \sim N(0,1)$ are the independent identically distributed (iid) random errors.
\end{itemize}
 
Models with more complex structure, such as multiplicative seasonal ARIMA models $$SARIMA(p,d,q)x(P,D,Q)_S$$ 
where S is the periodicity/frequency, D the seasonal differences; P auto-regressive and Q mean average seasonal components \cite{Tsay2010}, and dynamic regression \cite{harima}, are location-scale models with additional changes in the basic ARIMA structure. Let's say $Y = \{Y_t \ | \ Y_t \sim N(\mu_t,\sigma) \}$ follows an ARIMA(p,d,q) model as in $\eqref{eq:1}$, then a dynamic regression is just adding independent terms to the location parameter 
$$(1-B)^d Y_t = \mu_t + (1-B)^d X_t b$$
$X_t$ are the additional variables (\textit{no time dependence considered}), b are the regression parameters, and every variable in $X_t$ has the same differences as  in the ARIMA model. A GARCH model proposed by \cite{Bollersev1986} as a generalization of an ARCH \citep{engle1982}, the location-scale structure is easier to be noticed, but is fair to recall that the time dependency structure is for the scale parameter.
$$Y_t = \mu_t + \sigma_{t}\epsilon_t$$ 
$$\sigma_t^2 = \sigma_0 +\sum_{i=1}^s\alpha_i \epsilon_{t-i}^2 +\sum_{i=1}^k \beta_{i}\sigma^2_t. $$

In this model, the location parameter is constant in time, and $\sigma_0$ is the arch constant parameter\footnote{ \cite{fGarch} denoted $\sigma_0$ as $\alpha_0$ or $\omega_0$}. An additional variation of the garch model is the student-t innovations with unknown degrees of freedom. Which implies adding latent parameters to the GARCH structure:
$$Y_t = \mu +  \epsilon_t \left( \dfrac{v-2}{v}\sigma^2_{t}\lambda_t \right)^{1/2} $$  
where the unknown degrees of freedom have an inverse gamma distribution ($\lambda_t \sim IG(v/2,v/2)$) and the hyper-parameter $v$ is unknown. A further discussion is given by \cite{fonseca2019effects}.

\subsection{Prior Distribution}

By default, \pkg{varstan} declares weakly informative normal priors for every lagged parameter\footnote{Lagged parameters are the ones in the ARMA or GARCH components, in $\eqref{eq:1}$ the $\phi_i$'s are the lagged parameters of the auto-regressive part}. Other distributions can be chosen and declared to every parameter in a dynamic way before the estimation process starts. For simplicity, varstan provides \code{parameters()} and \code{distribution()} functions, that prints the defined parameters for a specific model, and prints the available prior distributions for a specific parameter respectively.\\

The priors distributions for the constant mean ($\mu_0$) and regression coefficients, can be chosen between, \textit{normal, t student, Cauchy, gamma, uniform, and beta}.  For the constant scale parameter ($\sigma_0$) a \textit{gamma, inverse gamma (IG), half normal, chi square, half t-student, or half Cauchy} distributions.\\

In SARIMA models, non stationary or explosive process \cite{shumway2010time} could cause divergences in \proglang{Stan}'s estimation procedure. To avoid this, the ARMA coefficients are restricted to a $\Phi = [-1,1]$ domain, and the available prior distributions  are \textit{uniform, normal, and beta}\footnote{ In SARIMA we define in $\Phi$ domain} \footnote{ If $\theta \sim beta(\alpha,\beta)$ in $[0,1]$ then $\theta_1 = 2(\theta-1) \sim beta(\alpha,\beta)$ in $[-1,1]$ }.\\ 

For VARMA models, the covariance matrix $\Sigma_0$ is factorized in terms of a correlation matrix $\Omega_0$ and a diagonal matrix that has  the standard deviations on the non zero values, through:
$$\Sigma_0 = D_{ }\Omega_0 D$$   
where $D = diag(\sigma_1,\sigma_2,\ldots , \sigma_d)$ denotes the diagonal matrix with $\sigma_i$ elements that accepts priors just like scale constant parameter ($\sigma_0$). Following the  Stan recommendations, for $\Omega_0$ a LKJ-correlation prior with parameter $\zeta > 0$ by \cite{LEWANDOWSKI20091989} is proposed. For a further discussion of why this option is better than a conjugated inverse Whishart distribution see \cite{brms2017} and \cite{Ramjini}. \\

The prior distribution for $\alpha$ and $\beta$ parameters in a GARCH model can be chosen between \textit{normal, uniform or beta distribution}, this is due to their similarity to an auto-regressive coefficient constrained in [0,1]. For the MGARCH parameters a \textit{normal, t-student, Cauchy, gamma, uniform or beta} distributions can be chosen. Finally, for the $v$ hyper-parameter in a GARCH model with unknown degree freedom innovations, a \textit{normal, inverse gamma, exponential, gamma, and a non-informative Jeffrey's prior} are available \cite{Fonseca2008}.

\section{Estimation process} \label{sec:estimation}

Just like \pkg{brms} \citep{brms2017} or \pkg{rstanarm} \citep{rstanarm} packages, \pkg{varstan} does not fit the model parameters itself. It only provides a \proglang{Stan} \citep{Stan} interface in \proglang{R} \citep{R} and works exclusively with the extended Hamiltonian Monte Carlo \cite{DUANE1987216}, No U-Turn Sampler algorithm of \cite{hoffman14a}. The main reasons for using this sampler are its fast convergences and its less correlated samples. \cite{brms2017} compares between Hamiltonian Monte Carlo and other MCMC methods, and \cite{betancourt2017} provides a conceptual introduction of the Hamiltonian Monte  Carlo. \\  

After fitting the model, \pkg{varstan} provides functions to extract the posterior residuals and fitted values, such as the predictive m-steps ahead and  predictive errors distribution. For model selection criteria, \pkg{varstan} provides posterior sample draws for the pointwise log-likelihood, Akaike Information Criteria (AIC), corrected AIC (AICc) and Bayesian Information Criteria (BIC) \cite{BIC2006}. For a better performance in the model selection,  an adaptation of the \code{bayes_factor} of the \pkg{bridgesampling} \citep{bridgsampling2020} package, the Bayesian leave one out (\code{loo}), and the Watanabe Akaike information criteria (\code{waic}) from the \pkg{loo} \citep{loo} package  are provided \cite{Vehtari2016} and \cite{bayesfactor}.\\

The \code{bayes_factor()} approximates the model's marginal likelihood using the \code{bridgesampling} algorithm, see \cite{gronau2017} for further detail. The \code{waic} proposed by \cite{watanabe} is an improvement of the Deviance information criteria (\code{DIC}) proposed by \cite{David}. The \code{loo} is asymptotically equivalent to the \code{waic} \cite{watanabe} and is usually preferred over the second one \cite{Vehtari2016}.\\

\subsection{Automatic order selection in arima models}

Selecting an adequate order in a seasonal ARIMA model might be considered a difficult task. In Stan, an incorrect order selection might be considered an ill model, producing multiple divergent transitions. Several procedures for automatic order selection have been proposed \cite{Tsay2010}, \cite{Hannan} and \cite{gomez}. A Bayesian version of \cite{forecast2020} algorithm implemented in their \pkg{forecast} \citep{Rob2007} package is proposed. This adaptation consists in proposing several models and select the "best" one using a simple criteria such as \code{AIC}, \code{BIC} or \code{loglik}. Finally, fit the selected model.\\

In the proposed function, the \code{BIC} is used as selection criteria for several reasons: it can be fast computed, and it is asymptotically equivalent to the \code{bayes_factor}. After a model is selected,  the function fits the model with default weak informative priors. Even so, a \code{BIC} is a poor criteria for model selection. This methodology usually selects a good initial model with a small amount of divergences (usually solved with more iterations) delivering acceptable results. For further reading and discussion see \cite{Rob2007}.

\section{Package structure and modeling procedure} \label{sec:software}

Similar to \pkg{brms} \citep{brms2017} and \pkg{rstanarm} \citep{rstanarm}, \pkg{varstan} is an \proglang{R} interface for \proglang{Stan}, therefore, the \pkg{rstan} \citep{rstan} package and a \proglang{C++} compiler is required, the \href{https://github.com/stan-dev/rstan/wiki/
	RStan-Getting-Started}{https://github.com/stan-dev/rstan/wiki/
	RStan-Getting-Started} vignette has a detailed explanation of how to install all the prerequisites in every operative system (\emph{Windows, Mac and Linux}). We recommend to install \proglang{R}-4.0.0.0 version or hihger for avoiding compatibility problems. The current development version can be installed from \proglang{GitHub} using the next code:
\begin{CodeChunk}
\begin{CodeInput}
 R> if (!requireNamespace("remotes")) install.packages("remotes")
 
 R> remotes::install_github("asael697/varstan",dependencies = TRUE)
\end{CodeInput}
\end{CodeChunk}

The \pkg{varstan} dynamic is different from other packages. First, the parameters are not fitted after a model is declared, this was considerate so the user could select the parameter priors in a dynamic way and call the sampler with a satisfactory defined model. Second, all fitted model became a \code{varstan} \proglang{S3} class, the reason of this is to have available \code{summary}, \code{plot}, \code{diagnostic} and \code{predict} methods for every model 
regardless of its complexity. The procedure for a time series analysis with \pkg{varstan} is explained in the next steps:

\begin{enumerate}
	\item \emph{Prepare the data}: \pkg{varstan} package supports numeric, matrix and time series classes (\code{ts}).
	\item \emph{Select the model}: the \code{version()} function provides a list of the current models. Their interface is similar to \pkg{forecast} and \proglang{R}'s \pkg{stats} packages.These functions return a list with the necessary data to fit the model in \proglang{Stan}.
	
	\item \emph{Change the priors}: \pkg{varstan} package defines by default, weak informative priors. Functions like \code{set_prior()}, \code{get_prior} and \code{print()} aloud to change and check the models priors. 
	
	\emph{Other useful functions are \code{parameters()} that prints the parameter's names of a specified model, and \code{distribution()} prints the available prior distributions of a specified parameter.}
	
	\item \emph{Fit the model}: the \code{varstan()} function call \proglang{Stan}, and fit the defined model. Parameters like number of iterations and chains, warm-up,  and other Stan's control options are available. The \code{varstan()} class contains a \code{stanfit} object returned from the \pkg{rstan} package, that can be used for more complex Bayesian analysis.
		
	\item \emph{Check the model}: \code{summary()}, \code{plot()}/\code{autoplot()} and \code{extract_stan()} methods are available for model diagnostic and extract the parameter´s posterior chains.
	
	\emph{The \code{plot} and summary  methods will only provide general diagnostics and visualizations, for further analysis use the \pkg{bayesplot} \citep{bayesplot2019} package.}
	
	\item \emph{Select the model}: For multiple models, \pkg{varstan} provides \code{loglik()}, \code{posterior_residuals()}, \code{posterior_fit()}, \code{AIC()}, \code{BIC()}, \code{WAIC()}, \code{loo()} and \code{bayes_factor} functions for model selection criterias.
	\item \emph{Forecast}: the \code{posterior_predict()} function samples from the model's n-steps ahead predictive distribution.
\end{enumerate}

\section{Case study: Analyzing the monthly live birth in U.S. an example} \label{sec:example}

\begin{figure}[H]
	\centering
	\includegraphics[scale = 0.9]{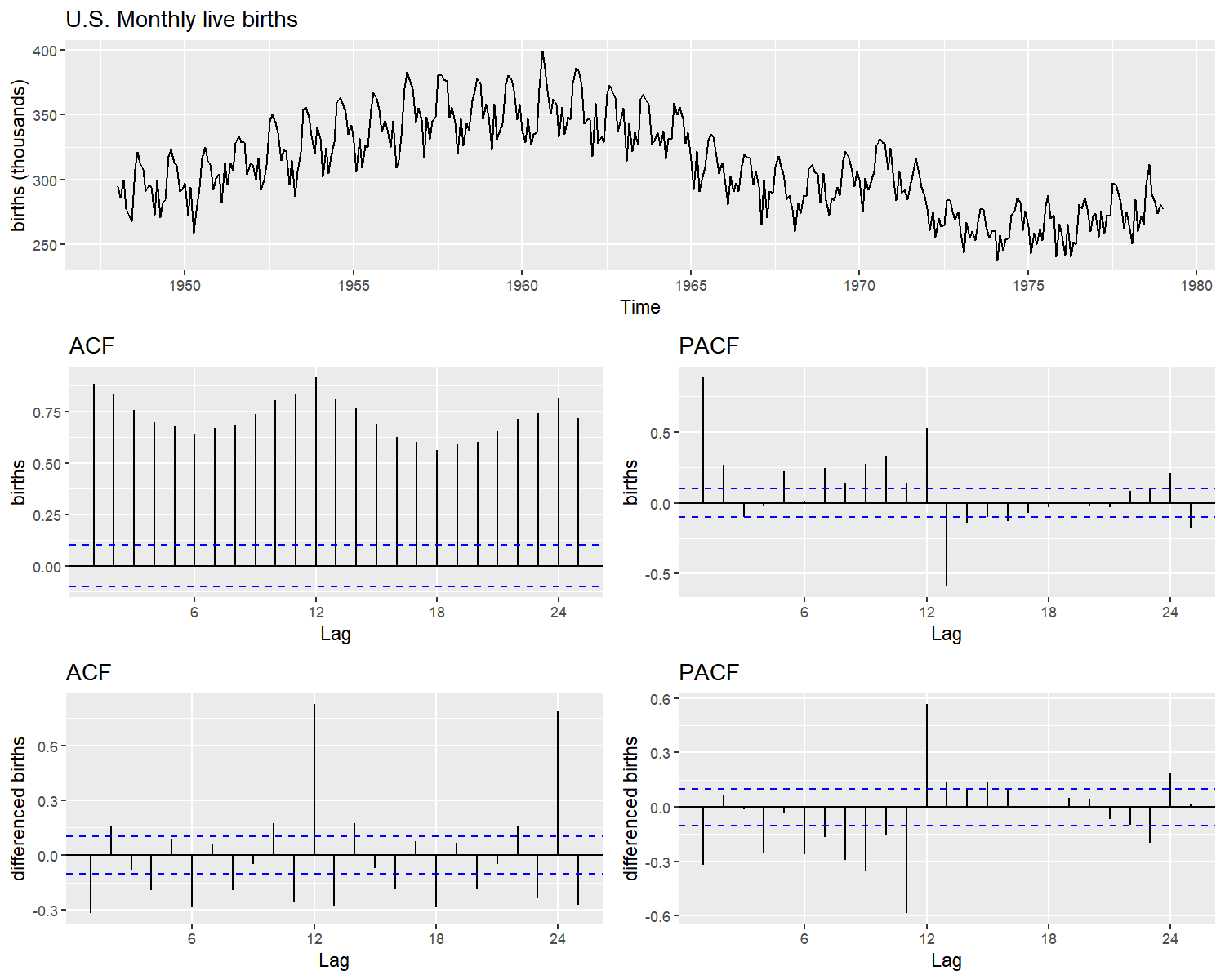}
	\caption[birth]{Monthly live birth U.S.A}
	\label{fig:fig1}
\end{figure}

As an example, a time series modeling for the monthly live births in the United States 1948-1979, published in \pkg{astsa} \citep{Stoffer2019} package is provided. In Figure \ref{fig:fig1}, the data has a seasonal behavior that repeats annually. The series \textit{waves} in the whole 40 years period (\textit{superior part}). In addition, the partial (\code{pacf}) and auto-correlation (\code{acf}) functions  are far from zero (\textit{middle part}), and have the same wave pattern as birth series. Indicating non stationary and a strong periodic behavior. After applying a difference to the data, the \code{acf} and \code{pacf} plots still have some non-zero values every 12 lags (\textit{inferior part}). \\ 

For start, a seasonal ARIMA model could give a good fit to the data. Following \cite{Tsay2010} recommendations for order selection using the auto-correlation functions, a \code{p = 1, d = 1, q = 1} and for the seasonal part \code{P = 1, D = 1,  Q = 1}. The model is defined in \pkg{varstan} as follows

\begin{CodeChunk}
\begin{CodeInput}
R> model1 = Sarima(birth,order = c(1,1,1),seasonal = c(1,1,1))
R> model1
\end{CodeInput}

\begin{CodeOutput}
y ~ Sarima(1,1,1)(1,1,1)[12] 
373 observations and 1 dimension 
Differences: 1 seasonal Differences: 1 
Current observations: 360 

Priors: 
Intercept:
mu0 ~ t (loc = 0 ,scl = 2.5 ,df = 6 )

Scale Parameter: 
sigma0 ~ half_t (loc = 0 ,scl = 1 ,df = 7 )

ar[ 1 ] ~ normal (mu =  0 , sd =  0.5 ) 
ma[ 1 ] ~ normal (mu =  0 , sd =  0.5 ) 

Seasonal Parameters: 
sar[ 1 ] ~ normal (mu =  0 , sd =  0.5 ) 
sma[ 1 ] ~ normal (mu =  0 , sd =  0.5 ) 
NULL
\end{CodeOutput}
\end{CodeChunk}

The function \code{Sarima} generates a Seasonal ARIMA model ready to be fitted in \proglang{Stan} \citep{Stan}. As the model is printed, all the important information is shown: the model to be fit, the total observations of the data, the seasonal period, the current observations that can be used after differences, and a list of priors for all the model's parameters. Using the information provided by the \code{acf-plot} in Figure \ref{fig:fig1} (\textit{middle right}), the partial auto-correlations are not that strong, and a normal distribution for the auto-regressive coefficient (\code{ar[1]}) could explore values close to 1 or -1, causing the prior to be too informative. Instead beta distribution in $[-1,1]$\footnote{ If $\theta \sim beta(\alpha,\beta)$ in $[0,1]$ then $\theta_1 = 2(\theta-1) \sim beta(\alpha,\beta)$ in $[-1,1]$ } centered at zero, could be a more proper prior. With the functions \code{set_prior()} and \code{get_prior()} any change is automatically updated and checked.

\begin{CodeChunk}
\begin{CodeInput}
R> model1 = set_prior(model = model1,dist = beta(2,2),par = "ar")
R> get_prior(model = model1,par = "ar")
\end{CodeInput}
	
\begin{CodeOutput}
 ar[ 1 ] ~ beta (form1 =  2 , form2 =  2 )
\end{CodeOutput}
\end{CodeChunk}

Now that the model and priors are defined, what follows is to fit the model using the \code{varstan()} function. One chain of 2,000 iterations and a warm-up of the first 1,000 chain's values is simulated.

\begin{CodeChunk}
\begin{CodeInput}
R> sfit1 = varstan(model = model1,chains = 1,iter = 2000,warmup = 1000)
R> sfit1
\end{CodeInput}
	
\begin{CodeOutput}	
y ~ Sarima(1,1,1)(1,1,1)[12] 
373 observations and 1 dimension 
Differences: 1 seasonal Differences: 1 
Current observations: 360 

mean     se       2.5%      97.5%      ess   Rhat
mu0        0.0061 0.0020     0.0020     0.0101 3941.739 1.0028
sigma0     7.3612 0.0043     7.3528     7.3697 4001.665 1.0000
phi       -0.2336 0.0014    -0.2362    -0.2309 3594.099 1.0007
theta      0.0692 0.0017     0.0658     0.0726 3808.853 1.0019
sphi      -0.0351 0.0015    -0.0381    -0.0321 3376.811 1.0033
stheta     0.6188 0.0017     0.6153     0.6222 3427.074 1.0048
loglik -1232.2519 0.0325 -1232.3157 -1232.1882 3198.671 1.0033

Samples were drawn using sampling(NUTS). For each parameter, ess
is the effective sample size, and Rhat is the potential
scale reduction factor on split chains (at convergence, Rhat = 1). 
\end{CodeOutput}
\end{CodeChunk}

\begin{figure}[H]
	\centering
	\includegraphics[scale =1]{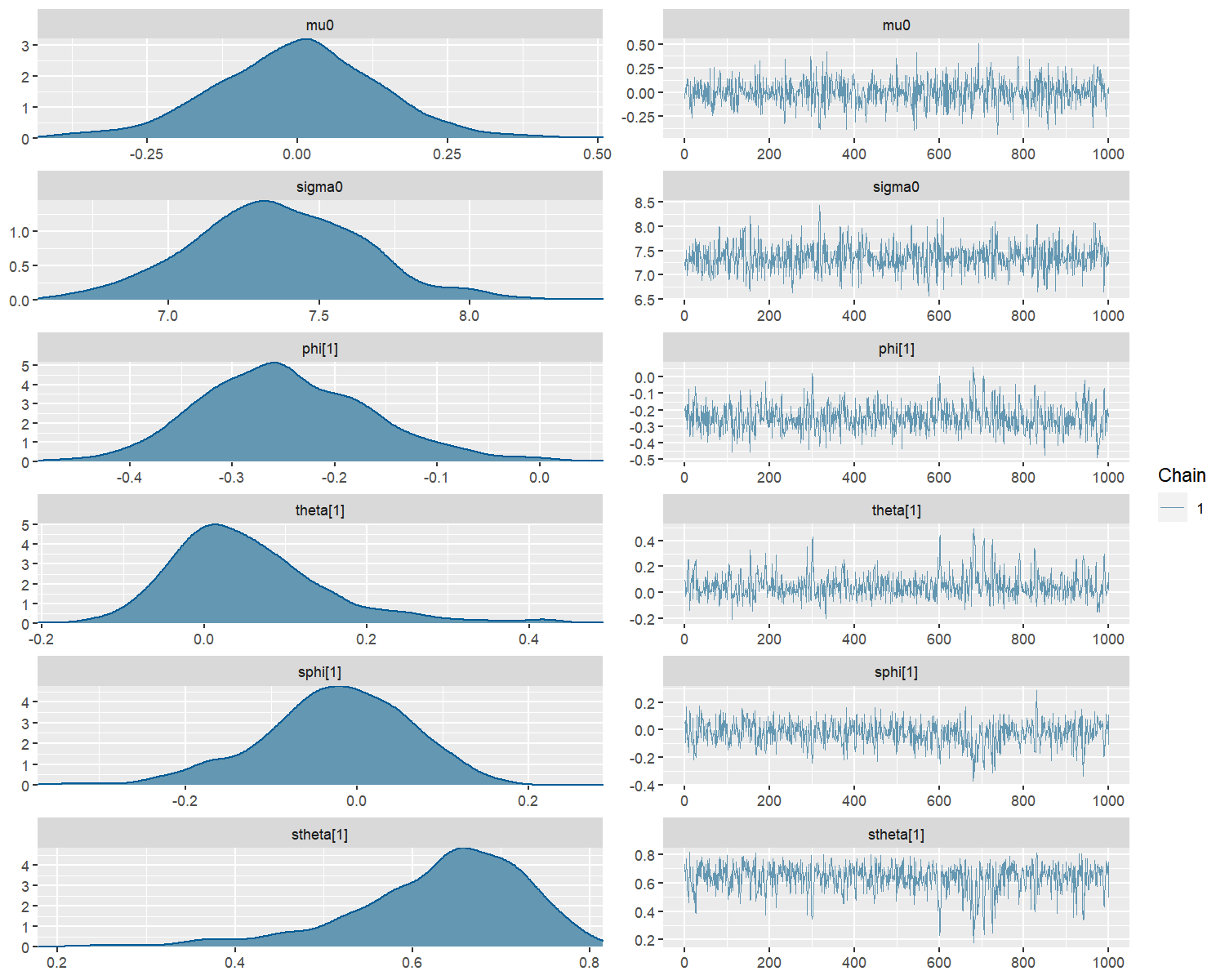}
	\caption[parameter]{Trace and density plots for all the fitted parameters}
	\label{fig:fig2}
\end{figure}

All fitted models are \pkg{varstan} objects, these  are S3 classes with the \code{stanfit} results provided by the \pkg{rstan} \citep{rstan} package, and other useful elements that make the modeling process easier. After fitting the proposed model, a visual diagnostic of parameters, check residuals and fitted values using the plot methods. On figure \ref{fig:fig2} trace and posterior density plots are illustrated for all the model parameters.

\begin{CodeChunk}
\begin{CodeInput}
R> plot(sfit1,type = "parameter")
\end{CodeInput}
\end{CodeChunk}

\begin{figure}[H]
	\centering
	\includegraphics[scale=1]{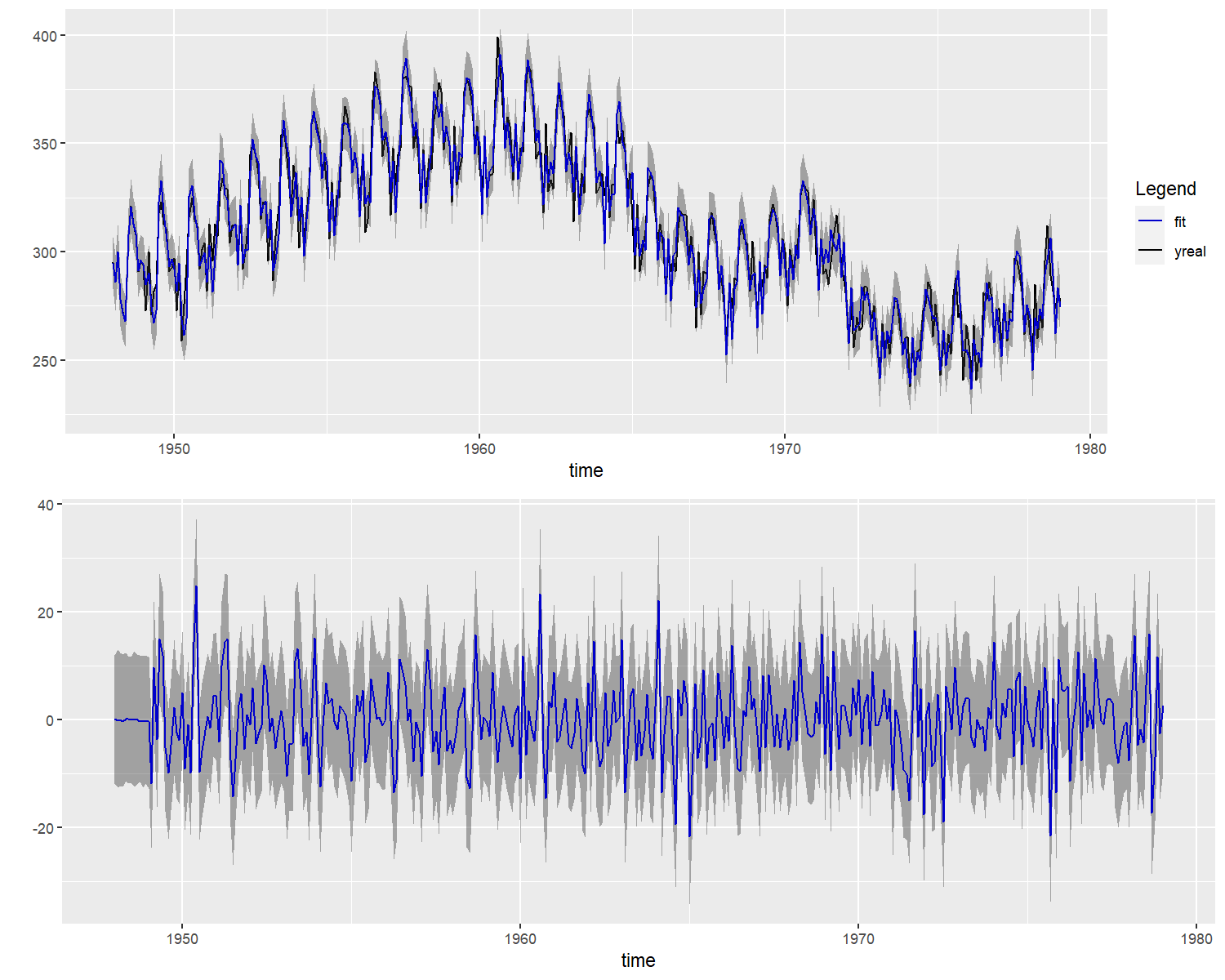}
	\caption[residual]{Posterior median residual plot}
	\label{fig:fig3}
\end{figure}

In figure \ref{fig:fig2}, all the chains appeared to be stationary, and the posteriors seems to have no multi-modal distributions. Indicating that all chains have mixed and converged. One useful way to assess models' fit, is by the residuals ($e_t = Y_t - \widehat{Y}_t$). The package provides the posterior sample of every residual, but checking all of them is an exhausting task. An alternative, is checking the process generated by the residuals posterior estimate. A white noise behavior indicates a good model fit. The model's residuals in figure \ref{fig:fig3}, seems to follow a random noise, the auto-correlation in \code{acf} plots quickly falls to zero, indicating an acceptable model fit.

\begin{CodeChunk}
\begin{CodeInput}
R> p1 = plot(sfit1,type = "residuals")
R> p2 = plot(sfit1)

R> grid.arrange(p2,p1,ncol = 1)
\end{CodeInput}
\end{CodeChunk}

Because of the sinusoidal pattern that birth series (\textit{figure }\ref{fig:fig1}) presents, a dynamic Harmonic regression (\textit{A fourier transform with arima structure for errors}) could also assess a good fit  \cite{harima}. To declare this model, varstan offers a similar declaration structure of \pkg{forecast} \citep{Rob2007} package. A harmonic regression with 4 fourier terms and ARIMA(1,1,1) residuals is declared and fitted to the birth data.

\begin{CodeChunk}
\begin{CodeInput}
R> model2 = Sarima(birth,order = c(1,1,1),xreg = fourier(birth,K = 2))
R> sfit2 = varstan(model = model2,chains = 1,iter = 2000,warmup = 1000)
R> sfit2
\end{CodeInput}

\begin{CodeOutput}	
y ~ Sarima(1,1,1).reg[4] 
373 observations and 1 dimension 
Differences: 1, seasonal Differences: 0 
Current observations: 372 

            mean     se       2.5%      97.5%       ess   Rhat
mu0       -0.0712 0.0068    -0.0846    -0.0578  939.6338 1.0032
sigma0    10.8085 0.0124    10.7841    10.8328  994.9722 1.0003
phi       -0.2705 0.0019    -0.2742    -0.2668 1004.4136 1.0006
theta      0.6242 0.0015     0.6212     0.6272  921.6062 0.9992
breg.1   -21.6318 0.0407   -21.7116   -21.5520  965.9837 1.0006
breg.2     0.6619 0.0305     0.6021     0.7217  976.7075 1.0002
breg.3     4.7937 0.0207     4.7531     4.8344 1079.1161 1.0003
breg.4    -5.3570 0.0249    -5.4059    -5.3082 1099.4993 1.0005
loglik -1415.3887 0.0697 -1415.5252 -1415.2521  936.0271 1.0005

Samples were drawn using sampling(NUTS). For each parameter, ess
is the effective sample size, and Rhat is the potential
scale reduction factor on split chains (at convergence, Rhat = 1). 
\end{CodeOutput}
\end{CodeChunk}
 
In this scenario both models suggest to be a good choice for birth series analysis. Even so, the harmonic regression fits more parameters. It is an obvious choice for birth's sinusoidal behavior. As an example of model selection criteria, the \code{bayes_factor} in logarithmic scale, that compares the models' marginal likelihoods is computed. Values above 6 (\textit{in logarithmic scale}) provides good evidence for selecting the first model. And for birth data, the seasonal arima model (\textit{model1}) is a better choice. 

\begin{CodeChunk}
\begin{CodeInput}
R> bayes_factor(x1 = sf1,x2 = sfit2,log = TRUE)
\end{CodeInput}
	
\begin{CodeOutput}	
Iteration: 1
Iteration: 2
Iteration: 3
Iteration: 4
Iteration: 5
Iteration: 6
Iteration: 1
Iteration: 2
Iteration: 3
Iteration: 4
Iteration: 5
Iteration: 6
Estimated log Bayes factor in favor of model1 over model2: 199.13745
\end{CodeOutput}
\end{CodeChunk}

Now, a comparison of the selected model (\textit{model1} $\sim$ \textit{Sarima(1,1,1)(1,1,1)[12]}) and the one given by the \code{auto.sarima()} function. For this purpose, a leave of one out cross validation \code{loo()} is used;, and both \code{looic} are compared with the \code{loo_compare()} function provided by the \pkg{loo} \citep{loo} package.

\begin{CodeChunk}
\begin{CodeInput}
R> sfit3 = auto.sarima(birth,chains = 1,iter = 4000)
R> sfit3
\end{CodeInput}
	
\begin{CodeOutput}	
y ~ Sarima(0,1,2)(1,1,1)[12] 
373 observations and 1 dimension 
Differences: 1 seasonal Diferences: 1 
Current observations: 360 

mean     se       2.5%      97.5%      ess   Rhat
mu0         0.0080 0.0018     0.0045     0.0116 2050.372 0.9997
sigma0      7.3517 0.0060     7.3399     7.3634 1991.938 1.0008
theta.1     0.3642 0.0013     0.3616     0.3668 1978.174 1.0006
theta.2     0.1358 0.0011     0.1336     0.1379 2023.769 1.0004
sphi       -0.2465 0.0016    -0.2496    -0.2433 2084.503 1.0005
stheta      0.3040 0.0017     0.3006     0.3073 2167.639 0.9995
loglik  -1231.7452 0.0395 -1231.8225 -1231.6679 1789.987 1.0009

Samples were drawn using sampling(NUTS). For each parameter, ess
is the effective sample size, and Rhat is the potential
scale reduction factor on split chains (at convergence, Rhat = 1).
\end{CodeOutput}
\end{CodeChunk}

Different from model1, the selected one does not contemplate an auto-regressive component, and use 2 mean average components instead.Now what proceeds is to estimate the \code{loo} for both models:

\begin{CodeChunk}
\begin{CodeInput}
R> loo1 = loo(sfit1)
R> loo3 = loo(sfit3)
		
R> lc = loo::loo_compare(loo1,loo3)
R> print(lc,simplify = FALSE)
\end{CodeInput}
	
\begin{CodeOutput}	
       elpd_diff se_diff elpd_loo se_elpd_loo p_loo   se_p_loo looic   se_looic
model2     0.0       0.0 -1235.4     15.4         7.1     0.8   2470    30.8 
model1    -0.8       5.8 -1236.2     15.6         7.8     0.9   2472    31.2 
\end{CodeOutput}
\end{CodeChunk}

\code{loo_compare()} prints first the best model. In this example is the one provided by the \code{auto.sarima()} function, where its \code{looic} is 2 units below model1. This \code{auto.sarima()} function is useful as starting point. But the reader is encouraged to test more complex models and priors that adjust to the initial beliefs.

\section*{Conclusions}

The paper gives a general overview of \pkg{varstan} package as a starting point of Bayesian time series analysis with structured models, and it offers a simple dynamic interface inspired in the classic functions provided by \pkg{forecast}, \pkg{astsa}, and \pkg{var} packages. The interface functions and prior flexibility that \pkg{varstan} offers, makes Bayesian analysis flexible as classic methods for structured linear time series models.  The package's goal is to provide a wide range of models with a prior selection flexibility. In a posterior version, non-linear models such as wackier GARCH variants, stochastic volatility, hidden Markov, state-space, and uni-variate Dynamic linear models will be included. Along with several improvements in the package's functionality.

\section*{Acknowledgments}

First of all, we would like to thank the Stan Development team  and the Stan forum's community for always being patience and helping fixing bugs and doubts in the package development. Furthermore, Paul Bürkner's encourage and advice made easier the development process, Enrique Rivera Gomez for his help in the article redaction and translation, Marco Cruz and Alicia Nieto-Reyes, for believe in this work. A special thanks to Mireya Matamoros for her help in every problem we have encountered so far.

%% -- Bibliography -------------------------------------------------------------
%% - References need to be provided in a .bib BibTeX database.
%% - All references should be made with \cite, \citet, \citep, \citealp etc.
%%   (and never hard-coded). See the FAQ for details.
%% - JSS-specific markup (\proglang, \pkg, \code) should be used in the .bib.
%% - Titles in the .bib should be in title case.
%% - DOIs should be included where available.

\bibliography{refs}

%% -- Appendix (if any) --------------------------------------------------------
%% - After the bibliography with page break.
%% - With proper section titles and _not_ just "Appendix".

%% -----------------------------------------------------------------------------

\end{document}